\documentclass[twocolumn,showpacs,preprintnumbers,amsmath,amssymb]{revtex4}
\usepackage{epsfig}
\usepackage{bbm}
\newcommand{\bm}[1]{ \mbox{\boldmath $#1$}  }

\begin{document}

\title{Relative production rates of $^{6}$He, $^{9}$Be, $^{12}$C in 
astrophysical environments}

\author{R. de Diego and E. Garrido}  
\affiliation{ Instituto de Estructura de la Materia, CSIC,  
Serrano 123, E-28006 Madrid, Spain } 

\author{D.V. Fedorov and A.S.~Jensen} 
\affiliation{ Department of Physics and Astronomy, 
         Aarhus University, DK-8000 Aarhus C, Denmark }

\date{\today}

\begin{abstract}
We assume an environment of neutrons and $\alpha$-particles of given
density and temperature where nuclear syntheses into $^{6}$He,
$^{9}$Be and $^{12}$C are possible.  We investigate the resulting
relative abundance as a function of density and temperature. When the
relative abundance of $\alpha$-particles $Y_{\alpha}$ is between $0.2$
and $0.9$, or larger than $0.9$, the largest production is $^{9}$Be or
$^{12}$C, respectively.  When $Y_{\alpha}<0.2$ $^{6}$He is mostly
frequently produced for temperatures above about $2$~GK whereas the
$^{9}$Be production dominates at smaller temperatures.
\end{abstract}



\maketitle

\paragraph*{Introduction.}

Formation of heavy elements must overcome the problem that all nuclear 
isotopes with mass numbers $5$ and $8$ are unstable \cite{apr05}. As a 
consequence, once the hydrogen fuel in a star is exhausted, the 
production of energy by formation of $^4$He stops and the temperature 
drops. The subsequent gravitational collapse increases the 
temperature, and the red giant phase, where helium is now the source 
of energy, begins. Due to the lack of neutrons in the core of the 
helium burning red giants, the $A$=5 and $A$=8 instability gaps have 
to be bridged by the triple-alpha reaction 
$\alpha+\alpha+\alpha\rightarrow \mbox{$^{12}$C}+\gamma$, which is the 
most relevant one in stars at the helium burning stage. 
 
Nevertheless, at conditions of high alpha and neutron densities, the 
reactions $\alpha+n+n\rightarrow \mbox{$^6$He}+\gamma$ and 
$\alpha+\alpha+n\rightarrow \mbox{$^9$Be}+\gamma$ 
are also possible, and they will play a role in bridging the $A=5,8$ gaps.  
As described in \cite{apr05,mey92,gor95}, such a scenario can appear 
at the early Big Bang stages or in the nucleosynthesis related to the 
type II supernova shock front. In both cases, temperatures are estimated 
to be of about 7 to 10 GK (1 GK=$10^9$ K). 
 
The dominating process is in any case the electromagnetic radiative recombination of  
the three particles from continuum to bound state, except at very 
high densities where four-body recombination can compete favorably 
\cite{die10}. However, the three-body processes producing these nuclei differ 
from each other. In particular, the production of $^{6}$He and $^{9}$Be is 
dominated by dipole transitions, while the production of $^{12}$C is of 
quadrupole character.

The three-body processes described above are usually treated as 
two-step processes where the unstable isotopes $^{5}$He and $^{8}$Be 
first are created and second, before decaying, react with another 
neutron or $\alpha$-particle \cite{efr96,sum02,bar06}.  This 
approximation of two independent sequential processes does not provide 
an accurate description since the lifetimes of the intermediate 
configurations are comparable to, or shorter than, the reaction time 
of the last step of the process \cite{fyn09}.  This implies that the 
processes proceed through genuine three-body reactions. 
 
The approximations employed to overcome this difficulty are, almost 
always,  extensions of the bare two-step model where the finite 
lifetime, or width, of the resonances is incorporated into the 
description \cite{bay83}. The fact that this can be tremendously wrong was 
discussed recently in a full three-body calculation of the triple 
$\alpha$-process at very low energies. In \cite{oga09} an increase 
in the triple alpha reaction rate by about 20 orders of magnitude 
around 10$^7$ K compared with the rate in \cite{ang99}  was found. In 
\cite{dot09} it is shown that such an increase is incompatible with 
observations of extended red giant branches and He burning stars in 
old stellar systems.  The three-body 
character is in general crucial for resonance structure calculations 
\cite{fed96,alv07} and the related dynamic evolution describing the decay  
mechanism \cite{gar09,alv07,alv08}. 
 
The correct three-body calculations are technically difficult because 
they involve the continuum dynamics of three particles interacting via 
a mixture of short and long-range forces.  Only recently it has become 
feasible to perform general three-body computations with the correct 
boundary conditions at both small and large distances, and where no   
assumption is made about the capture mechanism (sequential or direct).  
 
The purpose of the present letter is to investigate the only possible three-particle reactions 
at conditions of high alpha and neutron density. The novelty of this work is that we 
are doing it using a three-body method that is not making any assumption about the reaction  
mechanism.
In this way, we shall investigate the  
relative abundances of the three nuclei, $^{6}$He, $^{9}$Be and $^{12}$C,  
created from neutrons and $\alpha$-particles at given densities and  
temperatures. The conditions under which the production of each nucleus is 
the dominant process will be established.

\paragraph*{Formulation.} 
 
We consider radiative capture of three particles, $(abc)$, into a 
bound nucleus $A$ of binding energy $B_A$, i.e. $a+b+c\rightarrow 
A+\gamma$. The particles are $\alpha$-particles and neutrons in 
various combinations.   
The corresponding 
reaction rate $R_{abc}(E)$ can be related to the inverse process of 
photodissociation. Eq.(20) in \cite{fow67} gives the ratio between the ``energy 
averaged" reaction rates for the processes $a+b+c\leftrightarrow A+d$. Making use of (32) in 
\cite{fow67} and specializing to the case in which $d$ is a photon, one can identify the 
relation:  
\begin{equation} \label{eq7} 
R_{abc}(E)=\frac{\hbar^3}{c^2} \frac{8\pi}{(\mu_x \mu_y)^{3/2}} 
\left( \frac{E_\gamma}{E} \right)^2 \frac{2 g_A}{g_a g_b g_c} 
\sigma_\gamma(E_\gamma) 
\end{equation} 
where $E=E_\gamma+B_A$ is the initial three-body kinetic energy, 
$E_\gamma$ is the photon energy, $\sigma_\gamma(E_\gamma)$ is the 
photo dissociation cross section of the $A$ nucleus, $c$ is the 
velocity of light, $g_i$ is the degeneracy of particle $i=a,b,c,A$, 
and $\mu_x$ and $\mu_y$ are the reduced masses of the systems related 
to the Jacobi coordinates, $(\bm{x},\bm{y})$, for the three-body 
system \cite{nie01}. 
 
The production rate $P$ for the capture reaction is obtained after 
averaging $R_{abc}(E)$ using the Maxwell-Boltzmann distribution as 
weighting function, and multiplying by the density $n_i$ of particles 
$a$, $b$, and $c$. This density is usually written as $n_i=\rho N_A 
X_i/A_i$, where $\rho$ is the density of the environment, $N_A$ is the 
Avogadros number, $A_i$ is the mass number of particle $i$, and $X_i = 
N_i M_i /(N_a M_a + N_b M_b + N_c M_c)$ is the mass abundance of 
nucleus $i$ expressed by the number of particles $N_i$ and their 
masses $M_i$ (see Eqs.(1) and (3) in (\cite{fow67}). It is also possible to use the relative 
abundance defined by $Y_i = N_i /(N_a + N_b + N_c)$. 
 
In this way the final expression for the production rate $P$ depends 
then on both, temperature ($T$) and mass density ($\rho$) of the 
environment, which can vary substantially in different scenarios, and 
it takes the form \cite{fow67}: 
\begin{eqnarray} 
\lefteqn{  \hspace*{-1cm} 
P_{abc}(\rho,T)=n_a n_b n_c \frac{\hbar^3}{c^2} \frac{8\pi}{(\mu_x \mu_y)^{3/2}} 
\frac{g_A}{g_a g_b g_c} e^{-\frac{B}{k_B T}}\times } \nonumber \\ && \times 
\frac{1}{(k_B T)^3} 
\int_{|B|}^{\infty} E_\gamma^2 \sigma_\gamma(E_\gamma)e^{-\frac{E_\gamma}{k_B T}} dE_\gamma. 
\label{eq8} 
\end{eqnarray} 
 
The photodissociation cross section for the inverse process 
$A+\gamma\rightarrow a+b+c$ can be expanded into electric and magnetic 
multipoles. In particular, the electric multipole contribution 
of order $\lambda$ has the form \cite{for03}: 
\begin{equation} 
\sigma^{(\lambda)}_\gamma(E_\gamma)=\frac{(2 \pi)^3 (\lambda+1)}{\lambda [(2 \lambda+1)!!]^2}  
\left(\frac{E_\gamma}{\hbar c}\right)^{2 \lambda-1} \frac{d{\cal B}}{dE} \;, 
\label{eq8b} 
\end{equation} 
where the strength function ${\cal B}$ is 
\begin{equation} 
   {\cal B}(E\lambda,n_0J_0 \rightarrow nJ) = \sum_{\mu M} 
  |\langle nJM|O_{\mu}^{\lambda}|n_0J_0M_0\rangle|^2, 
\label{eq8c} 
\end{equation} 
where $J_0$, $J$ and $M_0$, $M$ are the total angular momenta and 
their projections of the initial and final states, and all the other 
needed quantum numbers are collected into $n_0$ and $n$.  The electric 
multipole operator is given by: 
\begin{eqnarray} \label{eq9} 
 O_{\mu}^{\lambda} = \sum_{i=1}^3 z_i |\bm{r}_i - \bm{R}|^{\lambda} 
 Y_{\lambda, \mu}(\Omega_{y_i}) \;, 
\end{eqnarray} 
where $i$ runs over the three particles, and where we neglect 
contributions from intrinsic transitions within each of the three 
constituents \cite{rom08}. 
 
\paragraph*{Test of the method.}

The crucial strength function, ${\cal B}$, is computed by genuine 
three-body calculations of both the bound final state and the 
continuum initial states. We have used the hyperspherical adiabatic 
expansion method described in \cite{nie01}. The $n$-$n$, $\alpha$-$n$, 
and $\alpha$-$\alpha$ interactions are the ones given in 
\cite{alv07,gar02}. The basic procedure is computation of three-body 
states of given angular momentum and parity confined by box boundary 
conditions \cite{die07}. In this way the continuum spectrum is 
discretized.  The strength functions are then obtained 
for each discrete continuum state according to Eq.(\ref{eq8c}). The 
distribution $d{\cal B}/dE$ is built by use of the finite energy 
interval approximation, where the energy range is divided into bins, 
and all the discrete values of ${\cal B}$ falling into a given bin are 
added. Afterwards the points are connected by spline operations and 
the expressions (\ref{eq8b}) and (\ref{eq8}) are computed. 
 
\begin{figure}[!h] 
\vspace*{-0.0cm} 
{\psfig{figure=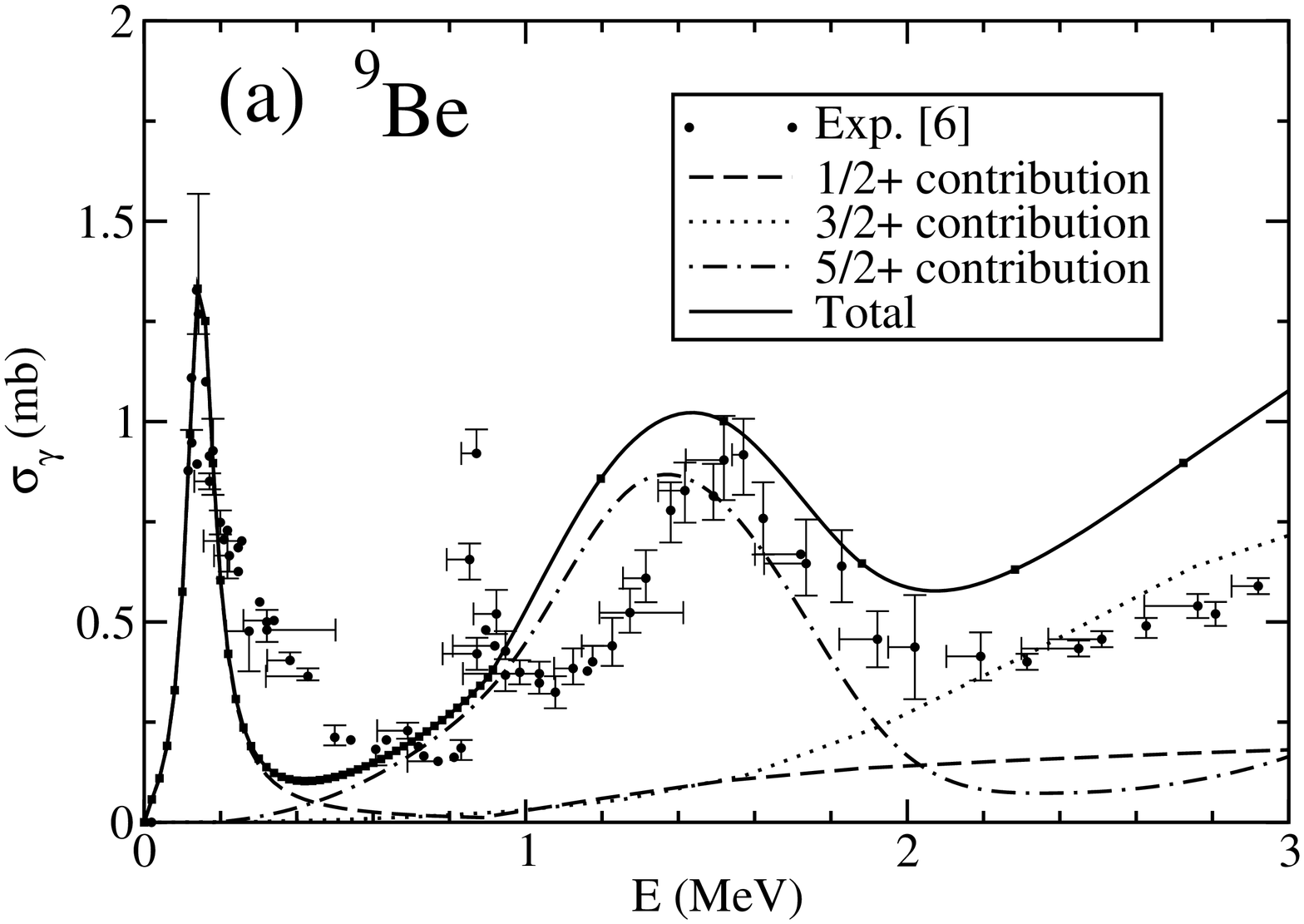,width=5.5cm,angle=270}} 
{\psfig{figure=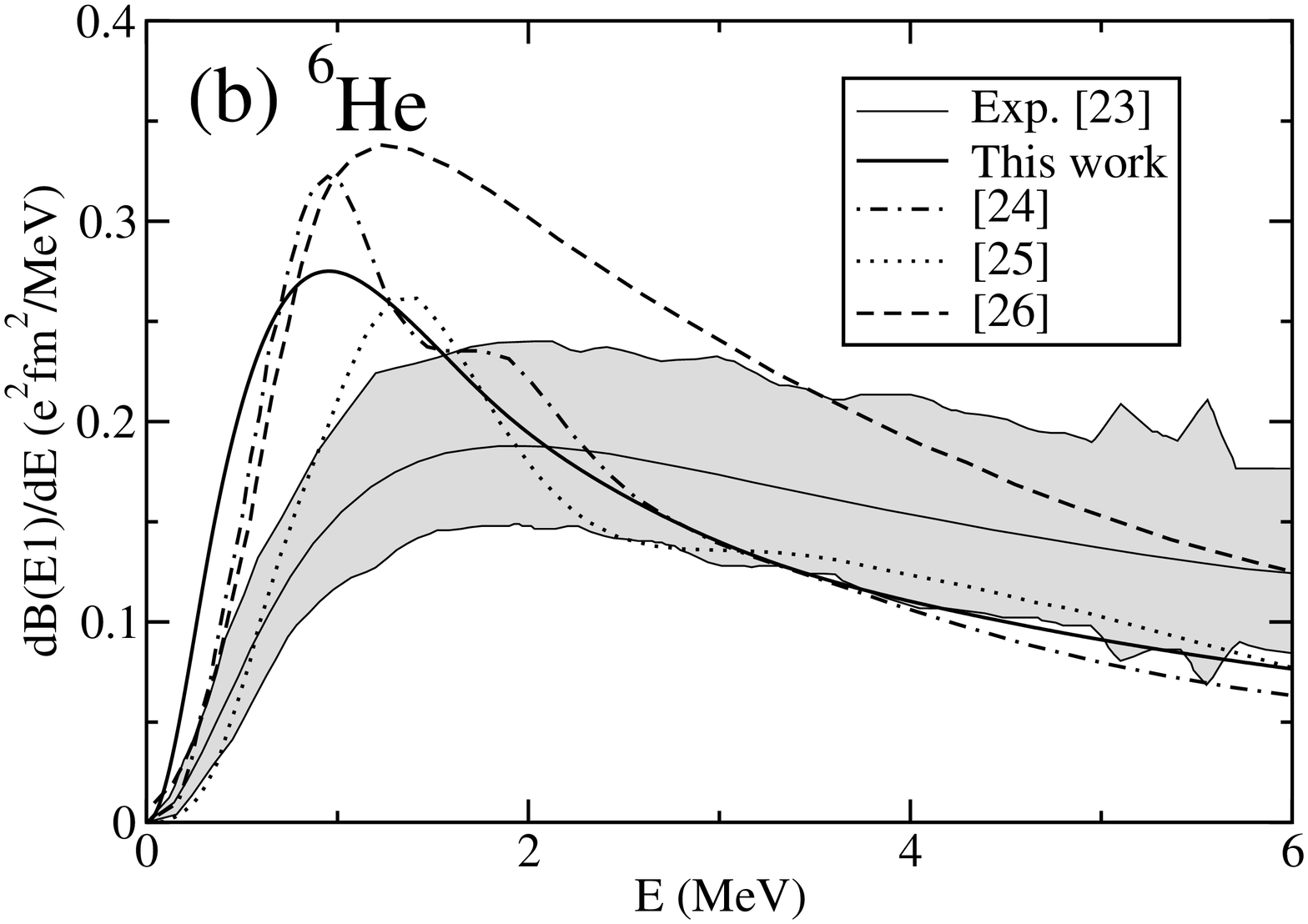,width=5.5cm,angle=270}} 
\caption{(a) Photodissociation cross section for $^9$Be. The total computed cross section is 
given by the solid line. The contributions from transitions to the 1/2$^+$, 
3/2$^+$, and 5/2$^+$ continuum three-body states are given by the dashed, dotted, and 
dot-dashed curves, respectively. The experimental data are from \cite{sum02}. (b) Comparison
between different computed dipole strength functions for $^6$He and the experimental data 
(shaded area) from \cite{aum99}.} 
\label{fig1} 
\end{figure} 
 
In the upper part of Fig.\ref{fig1} we show the computed photodissociation cross section of $^9$Be. 
The experimental data in the figure are from \cite{sum02}. The dashed, dotted, and dot-dashed curves give  
the contribution to the transition strength in Eq.(\ref{eq8c}), and therefore to the total cross  
section (solid line), 
from the $\alpha+\alpha+n$ continuum states with total spin and parity 1/2$^+$, 3/2$^+$, 
and 5/2$^+$, respectively (the ground state in $^9$Be has spin and parity 3/2$^-$). As seen in the 
figure, the agreement with the experiment is reasonably good, except for energies {above} 2 MeV, where 
the computed cross section clearly overestimates the experimental data. This is because  
the interaction used in the three-body calculation provides a $3/2^+$ resonance twice wider 
than the experimental one. The narrow peak at 0.9~MeV is a known $5/2^-$ resonance which can only be 
populated through the M1 transition which is neglected in this calculation. 

For $^6$He, our computed $dB/dE$ (thick solid line in the figure) agrees reasonably well with the previous 
calculations in \cite{cob97,dan98}. The one in \cite{myo01} gives clearly larger values, mainly for energies 
beyond 1 MeV. In any case, none of the calculations fits well the experimental data (thin solid line) 
for all the energies despite the large experimental error bars. All the calculations show a sharp enhancement 
in the low energy region which is not seen in the experiment. Further experimental data would help to 
clarify this issue. In similar measurements with $^{11}$Li, the peak in the experimental strength function 
moved to lower energies \cite{shi95,nak06} , and thus closer to our theoretical 
predictions \cite{gar02b}, after improvements in the detection of the low energy neutrons.
 
\paragraph*{Density dependence.} 
 
The density dependence of the production rates in Eq.(\ref{eq8}) is 
very simple for a given temperature. The basic reaction rate for only 
three particles has to be multiplied by the number of each species of 
particles. For a given total density $\rho$ found by adding neutron 
and $\alpha$-particle densities we can express the density dependence 
as $\rho^3 X_{\alpha}^n X_n^{3-n}$, where $X_{n}=1-X_{\alpha}$ and 
$X_{\alpha}$ are the mass fraction of neutrons and $\alpha$-particles. 
Then $n=1,2,3$ correspond to production of $^{6}$He, $^{9}$Be and 
$^{12}$C, respectively. 
 
When no $\alpha$-particles are present, $X_{\alpha}=Y_{\alpha}=0$, the  
production rates are all zero. When only 
$\alpha$-particles are present, $X_{\alpha}=Y_{\alpha}=1$, only $^{12}$C can be 
produced. The density dependence for production of $^{6}$He and 
$^{9}$Be are each others reflection in $X_{\alpha}=1/2$ but pushed 
towards smaller values as function of $Y_{\alpha}$.  The production of 
$^{12}$C increases monotonically as function of $X_{\alpha}$ and 
$Y_{\alpha}$.  
 
\begin{figure}[!h] 
{\psfig{figure=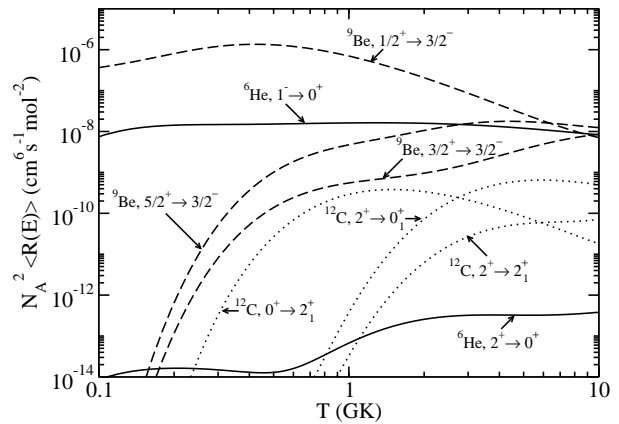,width=5.5cm,angle=270}} 
\caption{The rates as functions of temperature for producing  
$^6$He, $^9$Be and $^{12}$C in their respective ground states from 
various continuum states.  From $1^-$ and $2^+$ to $0^+$ for $^6$He 
(solid curves), from 1/2$^+$, 3/2$^+$, and 5/2$^+$ to 3/2$^-$ for 
$^9$Be (dashed curves), from $2^+$ to the $0^+$ ground state and from  
both $0^+$ and $2^+$ to $2^+$ excited bound state for $^{12}$C (dotted curves).} 
\label{fig2} 
\end{figure}

\paragraph*{Temperature dependence.} 
 
The temperature dependence is obtained by folding the calculated 
energy dependence with the Boltzmann distribution as seen formally in 
Eq.(\ref{eq8}) and shown numerically in Fig.\ref{fig2}. We show results 
from about 0.1 GK to 10 GK. The lower limit is where we run out of 
accuracy due to the
present discretization method, and the upper limit allows inclusion of
the relevant processes. For completeness we discuss the behavior in the
full interval. The 
three-body processes producing the three nuclei,$^{6}$He, $^{9}$Be, 
$^{12}$C, differ from each other.  Two neutrons and one 
$\alpha$-particle produce the $0^+$ ground state of $^{6}$He by 
electromagnetic recombination where one photon emerges with total 
angular momentum and parity $J^{\pi} = 1^-,2^+$. The dipole transition 
is strongly preferred in spite of the low-lying $2^+$ three-body 
resonance in $^{6}$He \cite{die07,die10}. 
 
One neutron and two $\alpha$-particles in continuum states of $J^{\pi} 
= 1/2^+,3/2^+,5/2^+$ combine directly by dipole transitions into the 
$3/2^-$ ground state of $^{9}$Be.  Resonances enhance the transitions 
in various energy ranges.  At low temperature transitions from the 
$1/2^+$ state dominate by several orders of magnitude but at $T\approx 
5$~GK the $5/2^+$ transition has almost reached the same value, see 
Fig.\ref{fig2}. This behavior is related to the corresponding 
resonances at excitation energies of $1.68$~MeV and $3.05$~MeV, 
respectively.  The $3/2^+$ resonance is located at a much higher 
energy and the corresponding transition always remains smaller. 
Furthermore, higher multipoles give insignificant contributions at 
these temperatures 
\cite{die10}. 
 
Three $\alpha$-particles can combine into the ground state of $^{12}$C 
through several paths. At low temperature the quadrupole transitions 
from $0^+$ dominates due to the Hoyle resonance, but the quadrupole 
transition from the $2^+$ continuum to the ground state increases and 
begin to dominate at $T\approx 2.5$~GK, see Fig.\ref{fig2}. The 
precise value depends on the energy of the lowest $2^+$ resonance 
which here was assumed to be $1.38$~MeV above the three-body threshold 
\cite{alv07}.  The remaining quadrupole transition from $2^+$ to $2^+$ 
is smaller for all temperatures. 
 
All these transitions are compared in Fig.\ref{fig2}. The production 
rate for $^9$Be is by far the largest at low temperatures but 
matched for $^6$He above temperatures of $T\approx 8$~GK. The rate for 
production $^{12}$C is much smaller in this temperature range. 
 
\begin{figure}[!h] 
{\psfig{figure=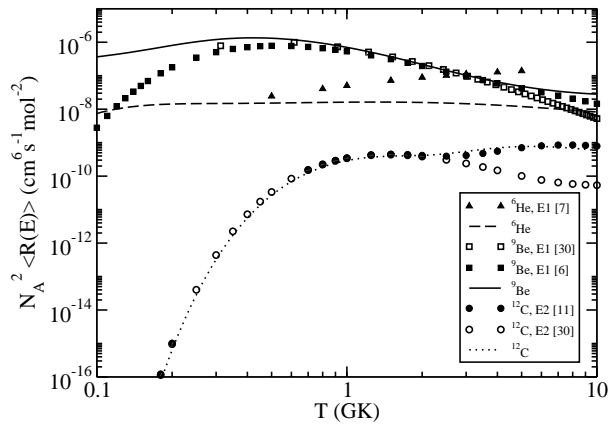,width=5.5cm,angle=270}} 
\caption{Comparison between the total computed rates for production of 
$^6$He (dashed line), $^9$Be, (solid line), and $^{12}$C (dotted line) 
and the ones calculated in \cite{bar06} for $^6$He, in \cite{sum02,fow88} for  
$^9$Be, and in \cite{ang99,fow88} for $^{12}$C.} 
\label{fig3} 
\end{figure} 
 
In Fig.\ref{fig3} the total reaction rates in Fig.\ref{fig2} are compared with 
previous calculations. The computed rates for $^6$He, $^9$Be, and $^{12}$C are given 
by the dashed, solid, and dotted curves, respectively. In \cite{gor95} the reaction 
rate for production of $^6$He is given by assuming a sequential capture process through 
the 3/2$^-$ resonance in $^5$He. This reaction rate is about three orders of magnitude smaller than  
our calculation, where the sequential capture assumption has not been made. In \cite{bar06} they  
give the same reaction rate but adding the contribution of a dineutron capture. This is shown 
by the closed triangles in the figure, which now an order of magnitude {\em above} our calculation.

For production of $^9$Be, our reaction rate agrees well with the ones in Refs.\cite{fow88} (open  
squares) and \cite{sum02} (closed squares) for temperatures above 1 GK. Below this temperature  
our reaction rate is clearly bigger. For $^{12}$C our computed rate agrees very well with the 
results given in \cite{ang99,fow88}, where a sequential capture through the narrow $0^+$ resonance 
in $^8$Be is assumed. The agreement with \cite{ang99} (closed circles) 
is particularly good. The deviations from \cite{fow88} (open circles) arise  
because we include the contribution from the $2^+\rightarrow 0^+$ transitions 
which is responsible for 
the enhancement of the tail for temperatures above 3 GK (see the dotted curves in Fig.\ref{fig2}). The good agreement with
\cite{ang99} does not imply that we disagree with the spectacular
results in \cite{oga09} where huge enhancements of the reaction rates
are reported for
energies around 0.01 GK.  We only discuss temperatures down to 0.1 GK
where the present method is applicable.  However, we see no indication
of such rate increase at very low temperature.
 

\paragraph*{Density-temperature dependence.}

The complicated temperature dependence in Fig.\ref{fig2} can now be 
combined with the simple density dependence, $n_a n_b n_c$, in Eq.(\ref{eq8}).  After 
adding up the different contributions for each of the three nuclei we 
find the creation probabilities as functions 
of temperature for three different relative fractions of the three 
nuclei. Creation of $^{12}$C only becomes competitive for any 
temperature when $Y_{\alpha}$ is close to unity, i.e. when essentially 
only $\alpha$-particles are present. 
 
The slightly increasing $^6$He curves cross the slightly decreasing 
$^9$Be curves while the $^{12}$C curves stay below for all temperatures below 
$10$~GK. Thus unless the relative $\alpha$-neutron abundance is extreme 
we get dominance of $^9$Be at small temperature and dominance of 
$^6$He at large temperature.  However, now it is important precisely 
which number of neutrons and $\alpha$-particles are available for the 
recombination process. When $Y_{\alpha}=0.1$ the dominance changes 
from $^9$Be to $^6$He at $T\approx 2.5$~GK. For $Y_{\alpha}$ 
values larger than 0.3 $^9$Be dominates for all temperatures.

\begin{figure}[!h] 
{\psfig{figure=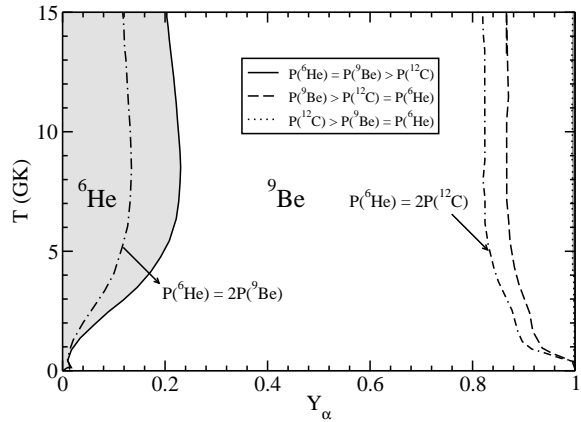,width=5.5cm,angle=270}} 
\caption{The phase diagram for producing $^6$He, $^9$Be and $^{12}$C 
in the $Y_{\alpha}$-temperature parameter space. The curves correspond 
to a constant ratio of production rates of two nuclei.} 
\label{fig5} 
\end{figure}

\paragraph*{Relative production.} 

The complete picture of the density-temperature
dependence of the three-body reaction rates for creating the three
nuclei is shown in Fig.\ref{fig5}.  All these nuclei can be destroyed in the
same environment but we consider here only the individual three-body
reactions creating them. Transfer reactions from $^6$He and $^9$Be using the
same basic ingredients of neutrons and alphas can also produce $^9$Be and
$^{12}$C.  However, we leave these four-body processes for future work,
they are beyond the scope of the present investigation.
For $Y_{\alpha}$ less than about $0.1$ and temperatures above $T\approx 
1-4$~GK the nucleus $^6$He is produced more than twice as often as 
$^9$Be.  As $Y_{\alpha}$ increases the relative $^9$Be production 
increases and becomes dominant for all temperatures when $Y_{\alpha}$ 
exceeds $0.2$. 
 
By further increase of $Y_{\alpha}$ the relative creation rate of 
$^{12}$C increases.  At $Y_{\alpha} \approx 0.82$ the $^9$Be 
production is still dominating when $^{12}$C is created with half the 
rate of $^{6}$He. At $Y_{\alpha} \approx 0.86$ the production rates 
for $^{12}$C and $^{6}$He are equal, but the production of $^9$Be 
still dominates.  Only when $Y_{\alpha}$ is larger than about $0.99$, 
where very few neutrons are present, the production rate of $^{12}$C 
exceeds the other rates.  These relative rates are very crudely 
independent of temperature except for very low $Y_{\alpha}$ values. 
Except for the small $Y_{\alpha}$ results, similar overall conclusions 
were obtained in previous investigations \cite{efr96,efr98,gor95}.

\paragraph*{Summary and conclusions.} 
 
We have compared calculated results of production rates of 
three-cluster nuclei, $^{6}$He, $^{9}$Be, $^{12}$C, consisting of 
neutrons and $\alpha$-particles.  The density and temperature 
dependence are investigated in a three-body model where the 
necessary continuum structures are treated by discretization in a large 
box. The two-body interactions are realistic and reproducing low-energy 
scattering properties.  The position of the lowest $2^+$ resonance in 
$^{12}$C is not precisely known but the results are rather insensitive 
to this energy. In general the results are very robust against 
variations within realistic limits of the interactions. 
 
The $^{12}$C production only dominates when the number of 
$\alpha$-particles is $100$ times larger than the number of neutrons. 
At lower relative $\alpha$-particle density the $^9$Be production is 
most frequent until $^6$He begins to dominate when the number of 
$\alpha$-particles is about $1/4$ of the number of neutrons.  An 
exception is low $\alpha$-particle density and temperatures below 
about $2$~GK where the $^9$Be rate is largest. Thus the actual 
temperature is important when only relatively few $\alpha$-particles 
are present. 
 
The full implications of the present results require detailed 
numerical investigations of the sequences of processes leading to 
formation of heavier nuclei. The initial conditions in these related 
chains of reactions should be our relative production rates. Whether 
the outcome is consistent with observations of nuclear abundances 
remains to be seen.  
 
\paragraph*{Acknowledgments.}  
This work was partly supported by funds provided by DGI of MEC (Spain) 
under contract No.  FIS2008-01301. One of us (R.D.) acknowledges 
support by a Ph.D. I3P grant from CSIC and the European 
Social Fund.

\end{document}